\documentclass[11pt]{article}
\usepackage[normalem]{ulem}
\usepackage{amsmath}
\usepackage{amsthm}
\usepackage{amsfonts}
\usepackage{bbm}
\usepackage{graphicx}
\usepackage{sectsty}
\RequirePackage{setspace}
\usepackage{appendix}
\usepackage{float}
\usepackage{url}
\usepackage{caption}
\usepackage{lineno}
\usepackage[round]{natbib}
\usepackage[table,xcdraw]{xcolor}
\usepackage{float}

\newcommand{\captionfonts}{\footnotesize}
\makeatletter 
\long\def\@makecaption#1#2{%
\vskip\abovecaptionskip
\sbox\@tempboxa{{\captionfonts #1: #2}}%
\ifdim \wd\@tempboxa >\hsize
{\captionfonts #1: #2\par}
\else
\hbox to\hsize{\hfil\box\@tempboxa\hfil}%
\fi
\vskip\belowcaptionskip}
\makeatother 

\begin{document}

\title{Are Colors Quanta of Light for Human Vision? A Quantum Cognition Study of Visual Perception}

\author{Jonito Aerts Argu\"elles \vspace{0.5 cm} \\ 
\normalsize\itshape
Center Leo Apostel for Interdisciplinary Studies, \\ \itshape Vrije Universiteit Brussel, 1050 Brussels, Belgium\vspace{0.5 cm} \\ 
\normalsize
E-Mail: \url{diederik.johannes.aerts@vub.be}
}

\date{}

\maketitle

\begin{abstract}
\noindent 
We show that colors are light quanta for human visual perception in a similar way as photons are light quanta for physical measurements of light waves. Our result relies on the identification in the quantum measurement process itself of the warping mechanism which is characteristic of human perception.
This warping mechanism makes stimuli classified into the same category perceived as more similar, while stimuli classified into different categories are perceived as more different.
In the quantum measurement process, the warping takes place between the pure states, which play the role played for human perception by the stimuli, and the density states after decoherence, which play the role played for human perception by the percepts. We use the natural metric for pure states, namely the normalized Fubini Study metric, to measure distances between pure states, and the natural metric for density states, namely the normalized trace-class metric, to measure distances between density states. We then show that when pure states lie within a well-defined region surrounding an eigenstate, the quantum measurement, namely the process of decoherence, contracts the distance between these pure states, while the reverse happens for pure states lying in a well-defined region between two eigenstates, for which the quantum measurement causes a dilation. We elaborate as an example the situation of a two-dimensional quantum measurement described by the Bloch model and apply it to the situation of two colors `Light' and `Dark'. We argue that this analogy of warping, on the one hand in human perception and on the other hand in the quantum measurement process, makes colors to be quanta of light for human vision.
\end{abstract}

\section{Introduction \label{introduction}}
We will argue that colors are quanta of light for human vision in a similar way that photons are quanta of light for a physical measuring device. To gather evidence for this argument, we will draw on various results obtained over the years within the research field of quantum cognition \citep{aertsaerts1995,aertsgabora2005b,busemeyeretal2006,aerts2009a,aerts2009b,bruzagabora2009,aertssozzo2011,aertsetal2012,busemeyerbruza2012,havenkhrennikov2013,khrennikov2014,dallachiaraetal2015,pothosetal2015,blutnerbeimgraben2016,moreirawichert2016,yearsley2017,aertsarguelles2018,busemeyeretal2019,surovetal2019,aertsarguellessozzo2020,aertsbeltran2022a}, some of which also involve our own work
\citep{aertsarguelles2018,aertsarguellessozzo2020}. Nevertheless, we want to make this article understandable to those who have not studied the investigations on which it relies, so that, in addition to references to this research and more technical portions of the article, we will also represent in an intuitive clear way the core aspects in which we frame our claim. 

Our argument rests on a fundamental property of human perception, which is not limited specifically to human vision, but is present in all forms of human perception, and which has its origin in a phenomenon called `categorical perception' \citep{harnad1987,goldstonehendrickson2010,aertsaertsarguelles2022}. Categorical perception `introduces a warping' of the `stimuli' to the `percepts' in a very specific way. That is, stimuli belonging to the same category are perceived in a way more similar than that shows by a physical measurement of the stimuli. Additionally, for stimuli belonging to different categories, the reverse warping takes place, and they are perceived in a way more different than a physical measurement of the stimuli shows. There is a long history \citep{libermanetal1957,libermanetal1967,lane1965,eimasetal1971,lawrence1949,berlinkay1969,daviesetal1998,davidoff2001,regierkay2009,havywaxman2016,hessetal2009,disaetal2011,sidman1994,schustermanetal2000,rosch1973,collieretal1973} in which the phenomenon of categorical perception was identified incrementally, and in Section \ref{categoricalperceptionandcolors} we give a brief description of this genesis. 

In the reasoning we develop, we also rely on a detailed study of the quantum measurement process. 
We have shown in \citet{aertsaertsarguellessozzo2025} that the quantum measurement process involves a warping that is the same as the warping that occurs in categorical perception. We will then gather the arguments that allow us to consider colors as quanta of light for human vision in a similar way that photons are considered quanta of light for physical measurements. For simplicity, we introduce the mathematical elements for our analysis for the case of two colors that we call {\it Light} and {\it Dark}, mentioning that fundamentally the same analysis can be made for multiple colors with mathematical elements that then belong to a more than two-dimensional complex Hilbert space. No other specific difficulties appear in this more dimensional situation that are not encountered in the two-dimensional case, namely that of two colors and a two-dimensional Hilbert space, which we consider in detail in the present article. 

In Section \ref{categoricalperceptionandcolors} we give a brief overview of the phenomenon of categorical perception with specific attention to the case of colors. In Section \ref{quantummeasurement} we introduce, in a detailed and self-prescriptive way, so that no prior knowledge is necessary, the quantum measurement model.
In Section \ref{quantummeasurement}, we give an overview of what was shown in \citet{aertsaertsarguellessozzo2025} and present the way that the quantum measurement changes distances between states illustrating that this way contains exactly the warping associated with categorical perception. In other words, the bias caused by human perception is intrinsically contained in the structure of the quantum measurement process. In Section \ref{colorsquanta} we gather the elements of the previous sections to show our claim that we can consider colors as quanta of light for human vision. 

\section{Categorical Perception, Quantization and Colors \label{categoricalperceptionandcolors}} 
In this section we wish to introduce the phenomenon of categorical perception because the nature of this phenomenon, along with the details of the measurement model of quantum mechanics, which our next Section \ref{quantummeasurement} deals with, plays a fundamental role in our claim that colors are quanta of light for human vision. It is interesting to reflect for a moment on the emergence of what was later called categorical perception because it illustrates well the role that certain goals and hypotheses can play. 

Around 1950, research on speech perception came into focus as work was being done on a speech device, the `pattern playback machine', which was intended to make it possible to automatically convert texts into spoken form so that, for example, blind people could read with it. This required analyzing speech very thoroughly, and it is within this setting that Alvin Liberman identified the phenomenon that would become known as categorical perception. More specifically, he noticed that when generating a continuum of evenly distributed consonant-vowel syllables with endpoints reliably identified as `b', `d' and `g', there is a point of rapid decrease in the probability of hearing the sound as `b' to hearing it as `d'. Later, there is a rapid switch from `d' to `g' \citep{libermanetal1957}. Liberman formulated an original hypothesis by which he wished to explain why people perceive an abrupt change between `b' and `p' in the way speech sounds are heard in contrast to what happens with a synthetic morphing device that produces the sounds with a continuous transition. His hypothesis was that this phenomenon is due to a limitation of the human speech apparatus, which, due to the muscular nature of its construction, would be unable to produce continuous transitions. Because of the way people produce these sounds as they speak, people's natural vocal apparatus would be unable to pronounce anything between `b' and `p'. So when someone hears a sound from the synthetic morphing device, that person tries to compare that sound with what he or she would have to do with his or her voice device to produce this sound. Since a human tuning device can only produce `b' or `p', all continuous synthetic stimuli will be perceived as `b' or `p', whichever is closest. 

The hypothesis was also the basis of what is now called the `motor theory of speech perception', which assumes that people perceive spoken words by recognizing the gestures in the speech channel used to pronounce them, rather than by identifying the sound patterns that produce the speech \citep{libermanetal1967}. The theory came under fire when it was found that `identification' and `discrimination' of stimuli not at all associated with speech behave in a similar way to stimuli associated with speech when measured in a similar manner \citep{lane1965}. Children, even before they could speak, were also found to have the specific categorical perception effect associated with speech perception that had been identified in adults \citep{eimasetal1971}. 

So, step by step, it became clear that categorical perception was a much more general phenomenon than just associated with speech, and the connection was made with earlier findings having to do with the way stimuli are organized. In particular, Lawrence's experiments and his hypothesis of `acquired distinctiveness' revealed a phenomenon that turned out to be a very basic effect of perception. The hypothesis of acquired distinctiveness states that stimuli for which one is taught to give a different response become more distinctive, while stimuli for which one is taught to give the same response become more similar \citep{lawrence1949}. Both effects are at work in humans in a multitude of perceptions, stimuli that fall within the same category are perceived as more similar, while stimuli that fall into different categories are perceived as more different. What happens with `colors' is a good example of the phenomenon of categorical perception. We see a discrete set of colors while from physical reality there is a continuum of different frequencies presented to us as stimuli. The warping effect of the categorical perception of colors consists in two stimuli that both fall within the category of, for example, green to be perceived more equally than two stimuli, one of which falls within the category of green and the second within the category of blue, even if from a physical perspective both pairs of stimuli have the same difference in frequency. The cooperation of these two effects, a contraction within an existing category and a dilation between different categories, causes a clumping of colors, ultimately leading to the colors that we distinguish. 

Given the research on colors that we present in this article, it is appropriate to mention Eleanor Rosch's work on colors. In fact, it was that work that inspired her to propose the prototype theory of concepts, which remains one of the most important theories of concepts today \citep{rosch1973}. The basic idea of prototype theory is that there exists a central element for a concept, which we call the prototype, relative to which the exemplars of the concept can be placed within a graded structure. Rosch suggested the idea that would develop into the primary model for concepts by studying the categorical structure among the Dani for colors and basic shapes. The Dani are people who live in Papua-New Guinea, with the peculiarity that they have only two words to denote colors, one meaning {\it Bright} and the other {\it Dark} \footnote{As in other articles of our Brussels research group, we denote concepts, when they appear as subjects in the text, by writing them beginning with a capital letter and in italic.}. The Dani also do not have words in their language for basic shapes such as {\it Circle}, {\it Square} and {\it Triangle}. Rosch examined whether there was a difference in learning between two groups of Dani volunteers, one group learning colors and basic shapes, starting with stimuli that are prototype colors and prototype basic shapes, while the other group learned to start with stimuli that are different distortions of these prototypes. In a significant way, it was found that for both colors and basic shapes, learning was more qualitative for the group taught starting from the prototype stimuli. This evaluation of `more qualitative learning' took into account the three features of how this can be measured, namely the ease of learning sets of shape categories when a particular type was the prototype, by the ease of learning individual types within sets, and by rank order of rating types as the best example of categories, when the prototypes of both colors and shapes were the stimuli in the learning process. 

Mathematical prototype models based on fuzzy sets were developed for concepts and experimentally tested, and it appeared that the way in which by warping, i.e. contraction when stimuli fall into the same category and dilation when they fall into different categories, concepts emerge and grow from stimuli had finally found a form to understand what is taking place at a fundamental theoretical level \citep{collieretal1973,rosch1975,roschetal1976,smithmedin1981,medinetal1984,geeraerts2001,johansenkruschke2005}. 

However, a fundamentally not understood problem, even when good mathematical models existed by which a concept and its set of exemplars and features could be modeled according to the approach of prototype theory, was the description of the combination of two concepts. This problem was noted in a first publication in which the combination of the concept {\it Pet} with the concept {\it Fish} served as an example, and therefore the problem of combining two concepts is often called the `pet-fish problem', or also the `guppy effect', because {\it Guppy} was the exemplar used to illustrate what goes wrong with prototype theory when concepts are combined \citep{oshersonsmith1981}. The question posed was `How is it that {\it Guppy} is not a typical example of a {\it Pet}, nor a typical example of a {\it Fish}, but a very typical example of a {\it Pet-Fish}'. It is starting from this `guppy effect' that in our Brussels research group it was tried successfully whether a quantum formalism could be used to model the guppy effect mathematically, where then the unexpected relevance of {\it Guppy} as a typical exemplar would be explained as an `interference effect' between the concepts {\it Pet} and {\it Fish} when combined \citep{aertsgabora2005a,aertsgabora2005b}. These first models using the quantum formalism were further developed and refined in the following years \citep{aerts2009a,aerts2009b} and their relevance to quantum information science was demonstrated by also modeling data obtained from the World Wide Web in a way similar to a quantum model \citep{aertsetal2012}. 

Meanwhile, it became clear that there were many more similarities between the structure and dynamics to which concepts of human language are subject and the structure and dynamics of quantum entities as described by quantum mechanics. The well-studied phenomenon of quantum entanglement, for example, also occurs with concepts, it is possible to combine concepts in a very similar way as is the case with quantum entities, such that experiments on these combinations violate Bell's inequalities, which is the experimental test for the presence of entanglement \citep{aertssozzo2011,aertssozzo2014}.
Another more recent finding consists of identifying the statistical properties of text of stories, short stories, and stories of length of novels. It could be shown in a very convincing way that the statistics inherent in such texts are of the Bose-Einstein type, hence the same as the statistics of a class of quantum entities called bosons, such as photons. It could also be shown that it cannot be of Maxwell-Boltzmann type, as one would expect it to be, since Maxwell-Boltzmann is the common statistics for a collection of classical entities \citep{aertsbeltran2020}. It was also investigated and shown that regarding the thermodynamic properties of texts representing stories, it is the quantum mechanical von Neumann entropy that is relevant and not the classical Shannon entropy \citep{aertsbeltran2022a,aertsbeltran2022b}. But the most important finding of a recent nature as far as the present article is concerned is that related to the phenomenon of quantization. It became clear that the phenomenon of categorical perception triggers a dynamics that leads to the emergence of quanta \citep{aertsaertsarguelles2022}. In the present article, we will again focus on categorical perception and how it gives rise to the formation of quanta, and we will make use of the result obtained in \citet{aertsaertsarguellessozzo2025}, namely that the mechanism of categorical perception is intrinsically present in the structure of the quantum measurement itself. We will then investigate in what way colors can be considered to be quanta of light for human vision.

Rosch, during her work on colors, assumed that the way humans see colors is determined, probably genetically, at birth, also the color frequencies of the basic colors. More so, the situation with colors being as such was actually her inspiration for proposing the prototype theory for concepts. 
This also means that the so-called Sapir-Whorf hypothesis, namely that there is an influence of `how we name categories within a language' and `the perception of stimuli belonging to these categories', was believed to be not applicable to colors.
This was generally accepted in the aftermath of Rosch's work on colors. Colors were thought to not depend on what they were called in different languages.
Not only do most cultures divide the groups of colors similarly and give them separate names, but for the few cultures where this is not the case, the areas of compression and dilation were assumed to be the same. It was believed that we all see blue and green in the same way, with a blurred area in between, even if the naming is not the same \citep{berlinkay1969}. However, this view was challenged by studies that nevertheless identified effects of the words designating colors. Comparative research on colors between Setswana speakers, a Bantu language spoken by about 8.2 million people in South Africa, and English speakers found many similarities, but also identified relevant differences in terms of the Sapir-Whorf hypothesis \citep{daviesetal1998}. The speskers of Berinmo, an indigenous language in Papua New Guinea, have only one word, `nol', for what English speakers call green and blue. The difference they made compared to English speakers in color discrimination tasks with respect to the shades between green and blue was investigated and determined \citep{davidoff2001}. Later, evidence was also found that linguistic categories influence categorical perception primarily in the right visual field. Since the right visual field is controlled by the left hemisphere, this finding was explained by the fact that language skills are also located in the left hemisphere \citep{regierkay2009}. 

More recent experiments, meanwhile, have demonstrated very thoroughly that language and the names given have an influence on the categorization that takes place over very primitive visual perceptions. Nine-month-old infants were shown a continua of new creature-like objects. There was a learning phase in which the infants were shown that objects from one end of the perceptual continuum moved to the left and objects from the other end moved to the right. For one group of infants, the objects were always called by the same name, while for the other group of infants, two different names were used to call the objects depending on whether they belonged to one end or the other of the continuum. The test involved showing new objects from the same continuum to all infants and then seeing if there was a difference between the two groups. What was found is that infants in the one-name condition formed one overarching category and looked at new test objects in either place. Infants in the two-name condition distinguished two categories and correctly predicted the likely location of test objects even when they were near the poles or near the middle of the continuum \citep{havywaxman2016}.

However, the experiment with the nine-month-old children we describe above, which demonstrates the effective existence of a Sapir-Whorf effect with colors, is not in contradiction with the findings that inspired Rosch. The colors we customarily distinguish do exist at birth, but the way these colors will play a role dynamically in an individual person's life, for this more advanced function of what colors are, is where a Sapir-Whorf effect does play a specific role. As we shall see further in the course of this article, this is the situation that is captured if colors are considered to be quanta of light for human visual perception, indeed, it is then the quantum mechanical structures of superposition and entanglement that can account for these effects. 

In her experiments with the Dani, Rosch used the basic colors defined by Brent Berlin and Paul Kay in their authoritative work `Basic Color Terms: Their Universality and Evolution' \citep{berlinkay1969}, and we will use these colors as basic colors in this article as well. 
By the way, let us note that these are not the colors of the rainbow, \citet{berlinkay1969} suggested the following eleven colors as basic colors within the culture where English is used as a native language, {\it White}, {\it Black}, {\it Red}, {\it Yellow}, {\it Green}, {\it Blue}, {\it Brown}, {\it Purple}, {\it Pink}, {\it Orange} and {\it Gray}. Berlin and Kay's work studies basic color naming in different language regions and concludes that seven stages in naming these basic colors can be distinguished from an evolutionary perspective. In the first stage, only two colors are named, {\it White} (then also meaning {\it Light}) and {\it Black} (then also meaning {\it Dark}). In a second stage, where there are three names for basic colors, {\it Red} is systematically added. In the next two stages, {\it Yellow} and {\it Green} are introduced, sometimes {\it Yellow} first and sometimes {\it Green} first. In the next stage, {\it Blue} is added, and {\it Brown} joins the names of the basic colors in the next stage. We have then reached the seventh stage where there are eight to eleven names for basic colors, the names still added being {\it Purple}, {\it Pink}, {\it Orange} and {\it Gray}.

The color system used both by Berlin and Kay and by Rosch, and in which they thus define the eleven prototype colors belonging to the English language, is the Munsell color system, a three-dimensional system where the sizes of the three parameters, hue, chroma and value define each color. Berlin and Kay use their original nomenclature of `focus colors' for the prototype colors, and Rosch also uses this nomenclature in her earliest articles on colors \citep{roschheider1971,roschheider1972,mervisetal1975}. Since it was mainly her research on colors that made Rosch develop the prototype theory for concepts, she calls these focus colors in her later work the prototype colors, and we will use this terminology in our writing on colors. 

In the next section, we discuss the connection between categorical perception and quantum measurement using the results obtained in \citet{aertsaertsarguellessozzo2025}. Hence, we look in detail at the situation of a two-dimensional quantum entity and show how its measurement dynamics is the underlying ground for the warping mechanism at work in the phenomenon of categorical perception.

\section{Quantum Measurement and Categorical Perception \label{quantummeasurement}}
In this section we will mainly write down what was proved in \citet{aertsaertsarguellessozzo2025}, namely that the warping characteristic of the phenomenon of categorical perception is also present in the measurement process of quantum mechanics. 
The analysis we did in \citet{aertsaertsarguellessozzo2025} is very thorough and detailed, and we will not repeat it here in its entirety. On the other hand, we want to keep the present article self-contained, so we also do not want to mention only the results. More so, we want to reproduce the understanding of `why the warping characteristic of the phenomenon of categorical perception is also the warping present in the quantum measurement apparatus'. We can do this in a simplified way because an intermediate result has been proved in \citet{aertsaertsarguellessozzo2025} from which we can start. This allows us to represent the insightful elements of the analysis in \citet{aertsaertsarguellessozzo2025} by using the simple geometry of triangles in a circle in the Bloch representation of a quantum measurement. 

\begin{figure}
\begin{center}
\includegraphics[width=11cm]{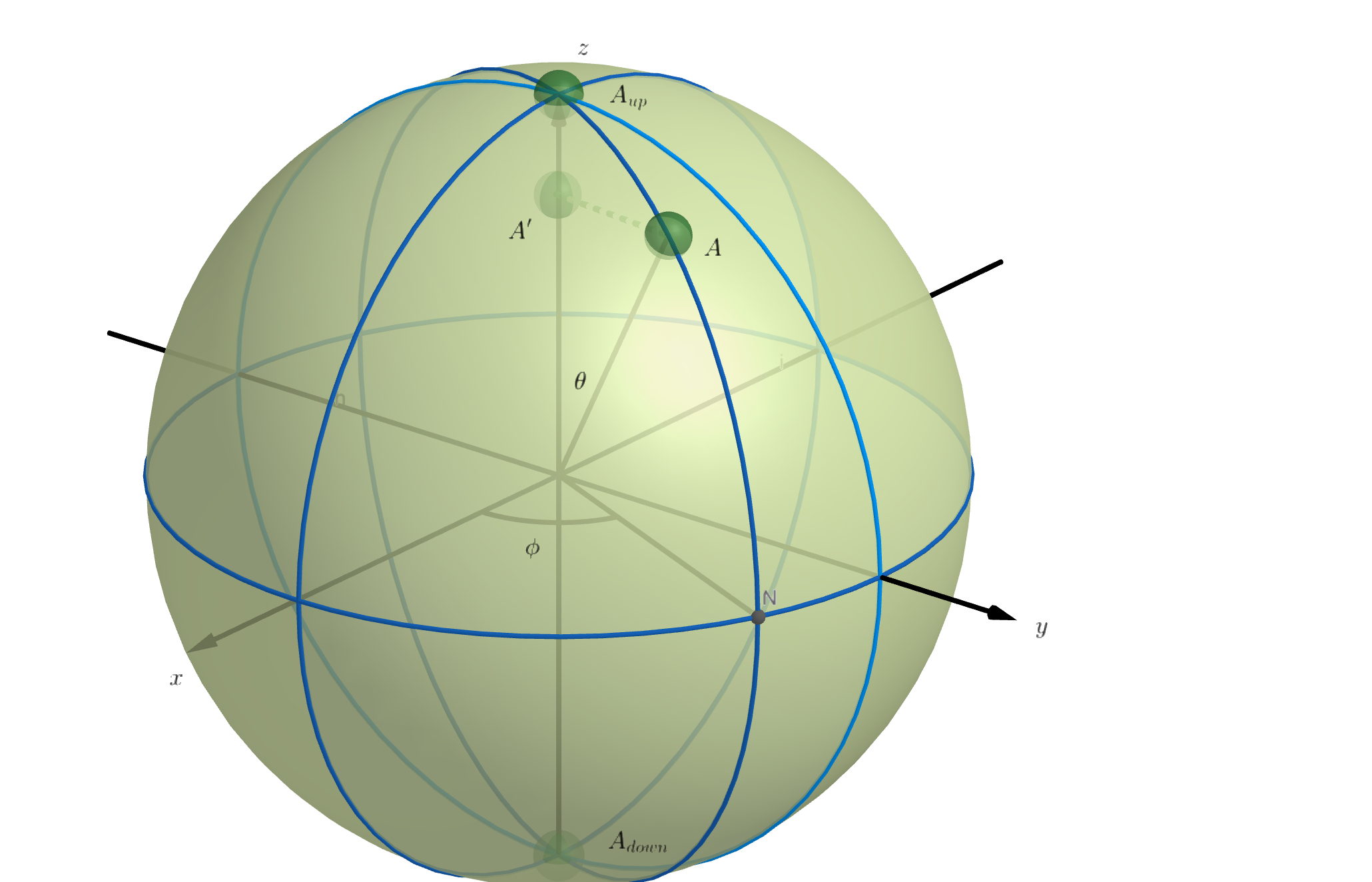}
\end{center}
\caption{A three-dimensional representation of the Bloch sphere. The quantum state of the considered quantum entity represented by point $A$ moves as a consequence of decoherence during a measurement on the line orthogonally to the diameter of the sphere to end up in the density state represented by the interior point $A'$ of the Bloch sphere and then collapses to one of the two eigen states of the measurement represented by points $A_{down}$ or $A_{up}$ respectively giving outcome `down' or `up'.}
\label{ElasticSphereModel3Dfigure}
\end{figure}

In Figure \ref{ElasticSphereModel3Dfigure}, we represent the Bloch sphere of a two-dimensional quantum entity, which is a sphere in three-dimensional Euclidean space with radius equal to 1. The Bloch sphere is the representation of the two-dimensional quantum entity in which we will prove our assertion. A two-dimensional quantum entity is also described in a two-dimensional complex Hilbert space, and in quantum computing it is called a `qubit'. Our assertion can also be for any larger dimension, and we will return to this explicitly in future work. Another well-known example of a two-dimensional quantum entity is the spin of a spin 1/2 quantum particle, and we keep this example in mind because it allows us to make some aspects of the proof in this section more demonstrative. 

In the Bloch representation, the pure states of the quantum entity are represented by the points on the surface of the Bloch sphere. The density states are represented by the interior points of the Bloch sphere. 
So, looking at Figure \ref{ElasticSphereModel3Dfigure}, the points $A$, $A_{down}$ and $A_{up}$, all points of the surface of the Bloch sphere, represent pure states of the quantum entity. The point $A'$, on the other hand, being an interior point of the Bloch sphere, represents a density state of the quantum entity. 
These four points also allow us to describe what happens during a measurement in quantum mechanics when this measurement is performed on the two-dimensional quantum entity we are considering. The point $A$ represents the state of the quantum entity before the measurement. For a two-dimensional quantum entity, each measurement always has two possible outcomes and when one of these outcomes takes place, the initial state of the quantum entity, represented by the point $A$ in our case, is changed into one of the two so-called eigenstates of the measurement, in Figure \ref{ElasticSphereModel3Dfigure}  we have represented these eigenstates by the points $A_{down}$ and $A_{up}$. So concretely, after performing the measurement, the quantum entity is either in the pure state $A_{down}$, and we say that the outcome took place which we call `down', or in the pure state $A_{up}$, and then we say that the outcome took place which we call `up'.
However, during the quantum measurement, the state of the quantum entity moves along a trajectory in the Bloch sphere. In a first stage, a process of decoherence takes place, the pure state in point $A$ falls orthogonally on the diameter connecting the two eigenstates and thus arrives at the final stage of decoherence in point $A'$.  We will come back to exactly what this process of decoherence means later. As we already mentioned, the point $A'$, being an interior point of the Bloch sphere, represents a density state of the quantum entity. From this point $A'$ the so-called quantum collapse takes place, the trajectory becomes discontinuous, and jumps either with outcome `down' to state $A_{down}$, or else with outcome `up' to $A_{up}$, again both pure states, thus the quantum coherence is restored.

The angles $\theta$ and $\phi$ are the angles of the so-called spherical coordinates. Specifically, $\theta$ is the angle that the line connecting the center of the Bloch sphere with the considered point $A$ of the surface of the Bloch sphere makes with the axis of the Bloch sphere, and thus varies between values $0$ and $\pi$, it is equal to $0$ if the point $A$ coincides with the North Pole of the Bloch sphere and it is called the polar angle (see Figure \ref{ElasticSphereModel3Dfigure}). Hence, it is equal to $\pi$ if it coincides with the South Pole \footnote{We denote the magnitude of angles by the unit of a radian, which makes 180 degrees equal to $\pi$ radians}. The angle $\phi$ is called the azimuthal angle, it is
the angle between a fixed axis in the horizontal plane of the equator and the line that this horizontal plane has in common with the vertical plane through the north and south poles and the point $A$ (see Figure \ref{ElasticSphereModel3Dfigure}). 
It varies between $0$ and $2\pi$. 

To verify that quantum measurement gives rise to the mechanism of categorical perception, we must be able to express for states how much they differ from each other. But before we get to that, we must identify which states play the role of stimuli and which states play the role of percepts in a quantum measurement process. The stimuli are represented by the states of the quantum entity independent of any measurement that would be going on, and hence they are the pure states. In Hilbert space, these pure states are represented by unit vectors of the Hilbert space. In the Bloch representation, they are represented by the points of the surface of the Bloch sphere. 
What are the percepts of a quantum measurement? 

We must dig deeper into the Hilbert space formalism than what shows us the representation by unit vectors of the pure states to see clearly in it, and the Bloch sphere representation is helpful in this regard. Indeed, the Hilbert space formalism, in addition to the vector space structure that allows us in a simple way to represent the pure states of a quantum entity, and to formulate the superposition principle, there is likewise the density operator formalism, carried by the same Hilbert space, but with its own structure that is further removed from the vector space structure, and is brought out in a less prominent way when quantum mechanics is discussed. Let us first note that the pure states do not coincide with the vectors of Hilbert space but with the unit vectors. In this sense, a superposition is not just a sum, but a sum followed by a renormalization, and is thus in its totality a nonlinear operation although directly using the linear operation, which the sum is. The density operator formalism uses the density operators of the Hilbert space. We will not further specify their mathematical structure at the level of the Hilbert space since they are represented in a more straightforward way by the interior points of the Bloch sphere. Indeed, all interior points of the Bloch sphere represent density states of the considered quantum entity.

It is important to note, and we cover this in depth and in great detail in \citet{aertsaertsarguellessozzo2025}, that there is no natural metric for the entire Bloch sphere. There does exist a natural metric on the interior points of the Bloch sphere, that is, on the density states, namely the trace distance. Suppose that $A$ and $B$ are two interior points of the Bloch sphere, the trace distance between $A$ and $B$ is then given by
\begin{eqnarray}
T(M(A), M(B)) &=& \frac{1}{2} Tr \left [\sqrt{(M(B)-M(A))^*(M(B)-M(A))} \right ]
\end{eqnarray}  
where $M(A)$ and $M(B)$ are the density matrices corresponding to the points $A$ and $B$.
It can be shown that for a two-dimensional quantum entity, i.e., a qubit, modeled with the Bloch sphere, the trace distance is equal to half the Euclidean distance in the three-dimensional Euclidean space of which the Bloch sphere is a part. It can also be shown that for two pure states $|\psi_1\rangle$ and $|\psi_2\rangle$, hence represented by two points of the surface of the Bloch sphere, the trace distance is given by
\begin{eqnarray} \label{tracepurestates}
   T(|\psi_1 \rangle \langle \psi_1 |, |\psi_2 \rangle \langle \psi_2 |) &=& \sqrt{1 - |\langle \psi_1 | \psi_2 \rangle |^2}
\end{eqnarray}
This general expression for the trace distance between pure states, makes it clear that the trace distance is not the natural metric to be used between pure states. Indeed, any effect due to quantum coherence disappears when using this metric, since in the general expression (\ref{tracepurestates}) the only reference to the states is the inner product between the two states, and hence quantum coherence effects are no longer present. 

There is also a purely geometric way by which we can see that the trace distance is unsuitable for measuring distances between pure states. Consider any two pure states, which gives us two points of the surface of the Bloch sphere. If we connect these two points with a line, we see that the points of the line different from these two points always lie in the interior of the Bloch sphere and thus correspond to density states. It is clear that defining a distance associated with a specific geometric entity, the straight line in the three-dimensional Euclidean space to which the Bloch sphere belongs, carries limitations when, as is the case with pure states, this distance is intrinsically intended for elements of a subset with a structure in which the straight line is not contained, and we mean here the spherical surface of the Bloch sphere to which the pure states are restricted.  

\begin{figure}[!ht]
\begin{center}
\includegraphics[width=8cm]{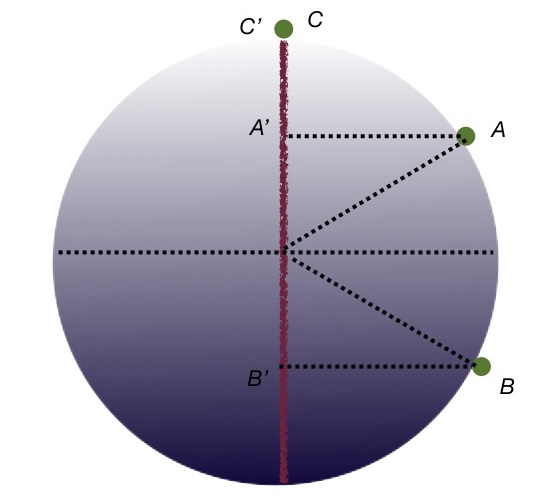}
\end{center}
\caption{We consider a situation where there are two names for colors, which we call {\it Light} and {\it Dark}, and wish to show that the quantum measurement model, which in this case represents a `qubit', incorporates the phenomenon of categorical perception. For this, we consider three pure states located in the Bloch representation in points $A$, $B$ and $C$, which represent stimuli associated with a quantum measurement on this qubit. With each of the three pure states corresponds a density state in which the qubit is located after the measurement, in the Bloch representation in respectively points $A'$, $B'$ and $C'$, and described by density matrices. The pure states in $A$ and $B$ belong to two different colors, {\it Light} and {\it Dark}, and lie at a distance 1/3 from each other. The density states corresponding to them, located in points $A'$ and $B'$, are at a distance 1/2 from each other. We see here the dilation mechanism of categorical reception at work, for percepts belonging to different categories, {\it Light} and {\it Dark}. The pure states in $C$ and $A$ belong to the same color, {\it Light}, and also lie at a distance 1/3 from each other. The density states corresponding to $C$ and $A$, located in points $C'$ and $A'$, are at a distance 1/4 from each other. We see here the contraction mechanism of categorical reception at work, for percepts belonging to the same category.}
\label{LightDarklabel}
\end{figure}

So we must ask ourselves at this point of our analysis which metric is the proper one to be used on pure quantum states. 
A natural notion of distance between pure states should only account for the pure states that are in-between the points on the Bloch sphere representing these states, the length of the circular arc connecting these points would be such a quantity. More precisely, if $\theta$ and $\alpha$ are the polar angles of the pure states $|\theta, \phi \rangle$ and $|\alpha, \phi \rangle$ (see Figure \ref{LightDarklabel}), the distance to be used to measure how much they are separated from each other, normalized to 1, would then be  
\begin{eqnarray}
    d_{pure}(|\theta, \phi \rangle, |\alpha, \phi \rangle) = \frac{1}{\pi} |\theta - \alpha|
\end{eqnarray} 
Note that it is sufficient to define the distance for two pure states with equal azimuthal angle $\phi$. Indeed, we can always choose a north pole and a south pole for any two pure states such that both states lie in the plane with equal azimuthal angle, so that then the arc of the circle through both points is determined by the difference of the polar angles of both states.

This distance also results from the angle between pure states calculated from their inner product of the Hilbert space, as was identified already in early twentieth century by two mathematicians Guido Fubini and Eduard Study to be the natural metric of the projective space of pure states, and it is now called the Fubini Study metric \citep{fubini1904,study1905}. 
For pure states $\psi_1$ and $\psi_2$ we have 
\begin{eqnarray}
    \gamma(\psi_1,\psi_2) = \arccos |\langle \psi_1 | \psi_2\rangle|
\end{eqnarray}
and normalized it coincides with the arc distance we introduced as natural distance on the pure states in the Bloch sphere. 

So, we will use two different notions of distance, one that considers the circular arc between two pure states, at the surface of the Bloch sphere, and the other one which considers the Euclidean distance between density states, inside the Bloch sphere. The normalized to 1 distance between two such density states 
will be given by (see Figure \ref{LightDarklabel}) 
\begin{eqnarray}
  d_{density}(\theta,\alpha) = \frac{1}{2} | \cos\theta - \cos\alpha |
\end{eqnarray}
Equipped with these two distances, let us now analyze how the phenomenon of categorical perception is naturally expressed in a quantum measurement process. 

As we mentioned, the points of the sphere's diameter between the South Pole and the North Pole are where the percepts lie for the measurement in question. 
To see how a quantum measurement brings about the warping effect of categorical perception, let us consider, to fix ideas, a situation where only two colors exist, {\it Light} and {\it Dark}, so we are precisely in a situation that can be described in a three-dimensional Bloch sphere. Note that Eleanor Rosch formulated the rationale for the prototype theory for concepts while teaching colors to a primitive community in Papua New Guinea, whose language, called Berinomo, has only two names for colors (Rosch 1973). 

Let us locate the first color, {\it Light}, at the North Pole of the Bloch sphere, and the second color, {\it Dark}, at its South Pole. At the equator, the transition from {\it Light} to {\it Dark} will then occur (see Figure \ref{LightDarklabel}). 
We introduce three pure states. The first one located in point $A$ (see Figure \ref{LightDarklabel}) with polar angle $\theta = \pi/3$, the second one located in point $B$ (see Figure \ref{LightDarklabel}) with polar angle $\theta = 2\pi/3$ and the third one located in the north pole (see Figure \ref{LightDarklabel}), (hence, this is the eigenstate describing {\it Light}, with polar angle $\theta=0$), assuming for simplicity that they all lie on the same plane, hence have the same asimuthal angle $\phi$. When a {\it Light}-{\it Dark} color-measurement is performed, the pure states deterministically transform into the fully decohered density states, obtained by plunging the associated points into the sphere, orthogonally with respect to the line subtended by the {\it Light} and {\it Dark} outcome states, which is the region of the percepts relative to this specific color measurement. This results for the pure states located in $A$, $B$ and $C$, respectively, in density states located in the points $A'$, $B'$ and $C'$.
 
If we consider the pure states located respectively in points $A$ and $B$, they are in the {\it Light} and {\it Dark} hemispheres of the Bloch sphere, hence they are two different colors, and we can now easily see that their transformation to the corresponding decohered density states, located respectively in points $A'$ and $B'$ (see Figure \ref{LightDarklabel}) and described by density matrices exhibits a warping that is a dilation. 
In fact, their distance is $1/3$, that is, one third of the maximal distance between two stimuli, while the associated percepts, corresponding to the decohered quantum states, have a distance of $1/2$, that is, one half of the maximal distance between two percepts. 

On the other hand, if we consider the pure states located respectively in points $C$ and $A$,  belonging to the same color, namely the color {\it Light}, the opposite warping occurs. Indeed, on the stimuli side, the distance is again $1/3$, whereas on the percepts side, the corresponding density states being located in points $C'$ and $A'$, respectiveky, and described by density matrices, we now have distance $1/4$. This means that a warping takes place, which is now a contraction. 

\begin{figure}[h!]
\begin{center}
\includegraphics[width=5cm]{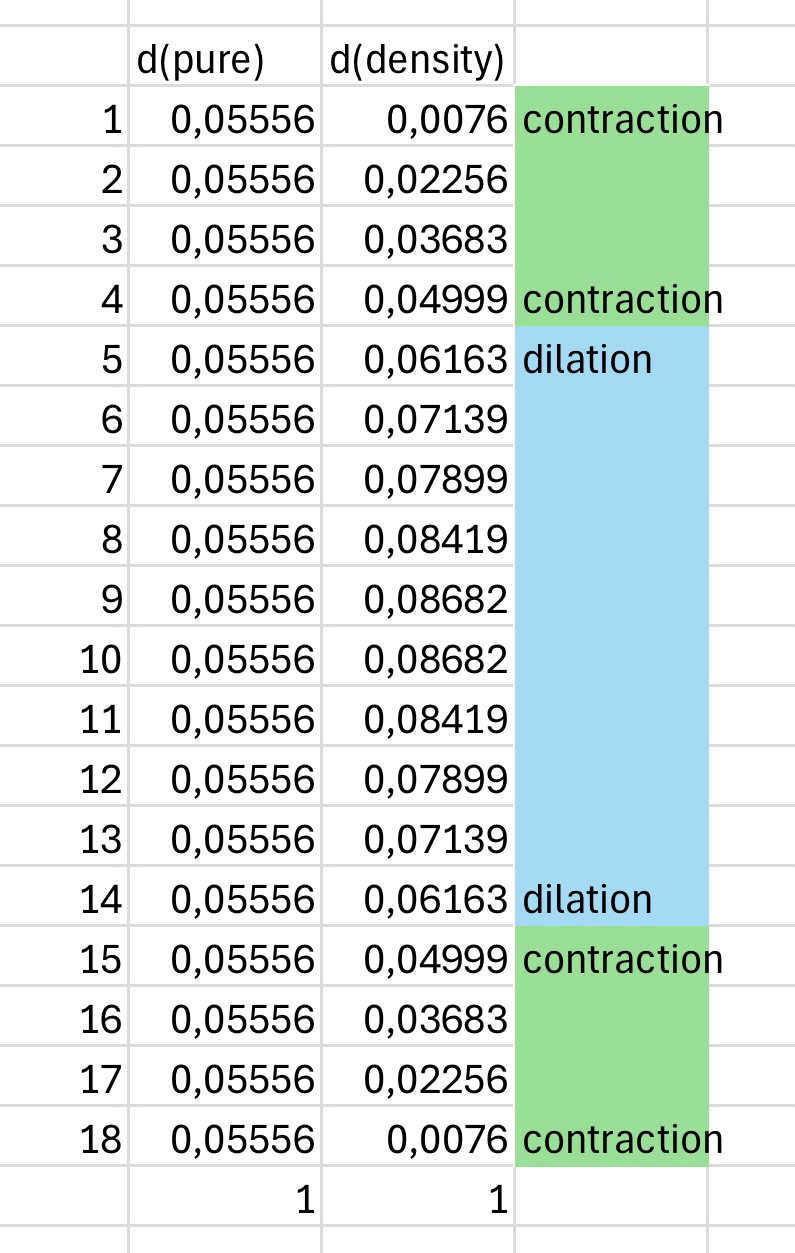}
\end{center}
\caption{A more detailed representation of the contraction or dilation that takes place in a quantum measurement. We have divided the $pi$ radians long circle arc connecting the North Pole with the South Pole in equal parts, each part spanning an angle of $0.05556 \cdot \pi$ radians or $10$ degrees. We thus determine $18$ points on the great circle arc connecting the North Pole and South Pole. In a region localized around the North Pole and a similar region localized around the South Pole, the decoherence of a quantum measurement causes the pure states associated with each of these points to be transformed into density states that are closer to each other and thus a contraction occurs. In the table, these cases are indicated in green. In a region located between the North Pole and the South Pole centered around the point $\frac{1}{2} \cdot \pi$ radians away from both, the decoherence of the quantum measurement causes a dilation. The pure states connecting to the points there are transformed into density states farther apart. In the table, these cases are indicated in blue.}
\label{Intervals}
\end{figure}

In Figure \ref{Intervals} we look in more detail at the effects of contraction or dilation, and have divided the $\pi$ radians or $180$ degrees circle arc connecting the North Pole with the South Pole into $18$ equal parts, each part spanning an angle of $0.05556 \cdot \pi$ radians or $10$ degrees. We thus determine 18 points on the great circle connecting the North Pole and the South Pole where each two consecutive points are $0.05556 \cdot \pi$ radians or $10$ degrees apart. In a region localized around the North Pole and a similar region localized around the South Pole, the decoherence of a quantum measurement causes the pure states associated with each of these points to be transformed into density states that are closer to each other, and thus a contraction occurs in these regions. In the table of Figure \ref{Intervals}, these cases are indicated in green. In a region located between the North pole and the South pole, centered around the point $\frac{1}{2} \pi$ radians away from both, the decoherence of quantum measurement causes dilation. The pure states connecting to the points there are transformed into density states farther apart. In the table of Figure \ref{Intervals}, these cases are indicated in blue. 
This shows how the phenomenon of categorical perception is built into the quantum measurement.

\section{Visual Perception, Colors and Quanta \label{colorsquanta}}

Let us analyze the events that correspond to this way of using the quantum measurement model to model the phenomenon of categorical perception of colors. And let us start from the situation where the stimulus, as measured by a physical measuring device that measures the frequency of electromagnetic radiation, is perceived by a human being, and from what we know takes place then. 

Suppose that the light being looked at has a frequency of 595 terahertz, given that 1 terahertz equals $10^{12}$ hertz, which corresponds to a wavelength of about 504 nanometers. 
\begin{figure}[h!]
\begin{center}
\includegraphics[width=5cm]{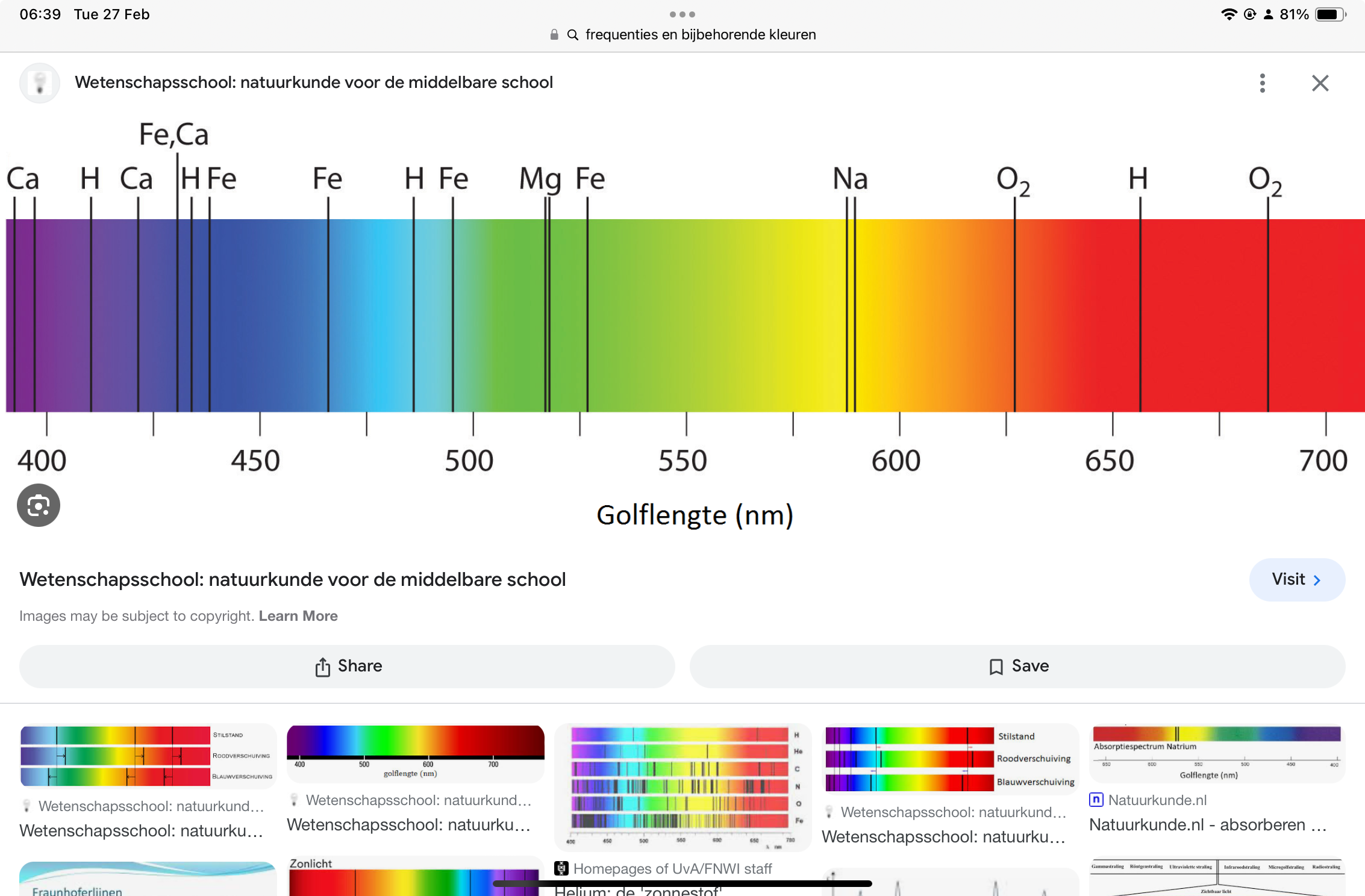}
\end{center}
\caption{A representation of the colors of visible light corresponding to the wavelengths of that light. Light with wavelength 504 nanometer has a color right between green and blue.}
\label{blue-greenlabel}
\end{figure}
If we consult one of the many color spectra of visible light available on the World Wide Web, we can see that the 504 nanometer wavelength corresponds to a color right between green and blue (see Figure \ref{blue-greenlabel}), so light of this frequency possesses a {\it Blue-Green} color, but it is not obvious whether when someone is asked to classify this light among one of the basic or prototype colors, whether that person will pick either {\it Blue} or {\it Green}. Each region that makes a transition from one basic color to another basic color contains a strip in which it is really not clear which of the adjacent basic colors will be chosen if such a choice is imposed, for example, in an experiment. 

That being said, we plan an experiment where we will show certain shades of color and on a Likert scale have participants choose to classify these shades of color with one of the basic colors. We will then investigate the extent to which we can model the shades of color in a complex Hilbert space as superpositions of the basis vectors, each representing one of the basic colors. Let us note that in our example of the shade {\it Blue-Green} carried by light of frequency 595 terahertz, this is a color carried by very pure light of exact frequency. We already noted that the prototype colors for the English language are `not' the colors of the rainbow, for example, the color {\it Brown} is not realized by light of a specific frequency. Colors of reflected light can represent very complex mixtures of different frequencies, just think of how white light is obtained by rotating a color disk rapidly. What happens there is very complex when the stimulus in question, the light shone from the rotating disk, is examined from physics. With respect to the nature of colors, what happens in our eye on the retina, but especially in our brain, and perhaps just as importantly, in and with our mind, is fundamental. Given the eleven prototype colors considered basic colors by \citet{berlinkay1969} and \citet{rosch1973}, for our planned experiment we will also consider these colors as basic colors. This means that the set of all superpositions is the set of elements, more precisely, the unit vectors, of an eleven-dimensional complex Hilbert space, with a basis of orthonormal vectors, each of which represents one of the considered basic colors. Although we can also explicitly build the Bloch representation for this situation of eleven basic colors, and we plan to do so after we have collected the data from our experiment, we will not become explicit about it in this article because the more than two-dimensional Bloch representations are more complicated, and we have focused, at least in part also for reasons of simplicity and intelligibility, but also because we wanted to turn to the fundamental in this article, to the situation of two colors, {\it Light} and {\it Dark}. After all, the fundamental, namely, the way in which stimuli, expectations, and measurements play a role in determining percepts, is equally and fully present in this situation of two colors.

So, let us return to the situation of two colors that we have called {\it Light} and {\it Dark}. 
Let it also be noted here that a restriction of vision to {\it Light} and {\it Dark}, and thus shades of gray, still leads to a powerful use of the sense of sight. In fact, there are several animals that see only shades of gray, and this group is called the `monochromatic mammals'. Several bats, rodents, and the common raccoon are among them. One common denominator among these mentioned species is that they are nocturnal animals, so their monochromatic vision gives them a special advantage at night. The gray shade that contains the most uncertainty regarding being classified as {\it Light} or {\it Dark} is the shade that lies at the equator of the Bloch sphere (see Figure \ref{LightDarklabel}). It is plausible to associate a probability equal to 1/2 with the point at the center of the line that runs between the north and south poles in the extended Bloch model. 

Are the probabilities proposed by the Bloch model also obtained in the limit of the fractions obtained for the outcomes when an experiment is conducted to measure them? Regarding this question, we should mention the result of research published in \citet{aertssassolidebianchi2015}. If we imagine different individuals and the way they will decide to classify a particular shade of gray as {\it Light} rather than {\it Dark} or vice versa, it is very plausible to assume that there will be a fair amount of individual differences. Some will still classify grays that are above the equator as {\it Dark} while others will still classify grays that are below the equator as {\it Light}. It is shown in \citet{aertssassolidebianchi2015} that the average over such individual differences tends to converge to a uniform distribution, which would then correspond to the quantum probabilities. This means that the quantum measurement model would be a kind of first-order approximation to more complex measurement situations, which would then also explain why in cases where control over such individual differences is absent, it leads to a good predictive model. It is beyond the scope of our current article to explain this complex situation of quantum as first-order approximation in more detail, and we refer the interested reader to \citet{aertssassolidebianchi2017} for a very detailed account of this situation.

Let us now come to our argument in connection with colors being quanta of light for human vision, and so for a raccoon these are only two colors {\it Light} and {\it Dark}, while for English-speaking humans they are eleven, {\it White}, {\it Black}, {\it Red}, {\it Yellow}, {\it Green}, {\it Blue}, {\it Brown}, {\it Purple}, {\it Pink}, {\it Orange} and {\it Gray}. Let us explain why these eleven colors are quanta of light for English-speaking humans. 
Consider the pet-fish problem we mentioned in Section \ref{categoricalperceptionandcolors}. In fact, in quantum cognition, such combinations of words are described by superpositions of the vectors that describe the participles of these combinations \citep{aertsgabora2005a,aertsgabora2005b,aertsetal2012}.

If we claim that our analysis makes it possible to consider colors as quanta for human vision, we must emphasize that there is an essential aspect of `human language' present. In fact, also in the anthropological research of \citet{berlinkay1969} the focus is on the existence of words that denote colors in a given language. That aspect was also already crucial in the situation of the Dani, as studied by \citet{roschheider1971,roschheider1972,rosch1973}, they possessed only two words denoting colors. In this sense, our observation that colors are quanta for the human vision of light is a special case of what is claimed in \citet{aertsbeltran2022a}, that `words are the quanta of human language'. Is it really the `words', which are the quanta of human language? Well, no, it is the underlying `concepts', as we made clear in \citet{aertsbeltran2022a}, and likewise in \citet{roschheider1971,roschheider1972,rosch1973}, it was all about the underlying concepts through which, incidentally, her research led to the introduction of the prototype theory for concepts. Is there no quantization before the emergence of language? This is an interesting question on which we want to dwell for a moment. There is no scientific consensus on the origin of human language, but one of the most plausible views postulates that the first languages consisted of gestures \citep{tomasello1996,pikamitani2006}. These gestures, with the disadvantage of making it difficult to communicate with someone in a more intimate way, evolved into movements accompanied by sounds more inward in the body, the mouth, lips, tongue, and larynx. Words and what we now commonly understand by language came much later, when more complex abstractions are communicated. It is our opinion that the phenomenon of categorical perception played a role from the very beginning, and probably thus then at the level of gestures, in instilling conceptual accumulations leading to clumping and finally quanta as a stabilization of this part of the process as a consequence of the contractions present on the stimuli and distinctions between the quanta as a consequence of the dilations. 

The time has come to clarify another aspect of our analysis. It rests on the consideration of the quantum measurement process that makes us interact with physical reality by means of measuring devices, on the one hand, and the consideration of human perception, which makes us interact with human conceptual reality by means of the human perceptual device, on the other hand, and then to consider both as special cases of a more general situation where measurement takes place on entities of a reality. 
In philosophy of mind, the notion of `qualia' was introduced to denote the subjective sensation of a human experience \citep{lewis1929}. Although there is no general consensus on this notion, neither among philosophers nor among scientists who study the human mind, such as neurologists, with some even doubting its existence, here we wish to use the notion in a specific way. In doing so, we do not pretend to resolve any controversy surrounding it, our intention is only to show that the notion of qualia can be used to shed additional light on the analysis we propose in this article about quantum measurement versus human perception. In this sense we claim that in a similar way that we can argue that `colors are quanta of light for human vision', we can argue that `photons are qualia for physical measuring devices that can identify them'.



\begin{thebibliography}{999}
\setlength{\itemsep}{-0.45 mm}




\bibitem[Aerts(2009a)]{aerts2009a} Aerts, D. (2009). Quantum structure in cognition. {\it Journal of Mathematical Psychology 53}, pp. 314-348. doi: \url{10.1016/j.jmp.2009.04.005}.

\bibitem[Aerts(2009b)]{aerts2009b} Aerts, D. (2009). Quantum particles as conceptual entities: A possible explanatory framework for quantum theory. {\it Foundations of Science 14}, pp. 361-411. doi: \url{10.1007/s10699-009-9166-y}.

\bibitem[Aerts \& Aerts(1995)]{aertsaerts1995} Aerts, D. and Aerts, S. (1995). Applications of quantum statistics in psychological studies of decision processes. {\it Foundations of Science 1}, pp. 85-97. doi: \url{10.1007/BF00208726}.

\bibitem[Aerts \& Aerts Argu\"elles(2022)]{aertsaertsarguelles2022} Aerts, D. and Aerts Argu\"elles, J. (2022). Human perception as a phenomenon of quantization. Entropy 24, 1207. doi: \url{10.3390/e24091207}.

\bibitem[Aerts et al., 2025]{aertsaertsarguellessozzo2025} Aerts, D., Aerts Argu\"elles, J. and Sozzo, S. (2025). Quantum measurement, entanglement and the warping of human perception.
In preparation.  

\bibitem[Aerts \& Beltran(2020)]{aertsbeltran2020} Aerts, D. and Beltran, L. (2020). Quantum Structure in Cognition: Human Language as a Boson Gas of Entangled Words. {\it Foundations of Science 25}, pp. 755-802. doi: \url{10.1007/s10699-019-09633-4}. 

\bibitem[Aerts \& Beltran(2022a)]{aertsbeltran2022a} Aerts, D. and Beltran, L. (2022a). Are words the quanta of human language? Extending the domain of quantum cognition. {\it Entropy 24}, 6. doi: \url{10.3390/e24010006}. 

\bibitem[Aerts \& Beltran(2022b)]{aertsbeltran2022b} Aerts, D. and Beltran, L. (2022b). A Planck Radiation and Quantization Scheme for Human Cognition and Language. Frontiers in Psychology 13, 850725. doi: \url{10.3389/fpsyg.2022.850725}.

\bibitem[Aerts et al.(2012)]{aertsetal2012} Aerts, D., Broekaert, J., Gabora, L. and Veloz, T. (2012). The guppy effect as interference. Quantum Interaction. {\it Lecture Notes in Computer Science 7620}, pp 36-47. doi: \url{10.1007/978-3-642-35659-9_4}.

\bibitem[Aerts \& Gabora(2005a)]{aertsgabora2005a} Aerts, D and Gabora, L. (2005a). A theory of concepts and their combinations I: The structure of the sets of contexts and properties.{\it Kybernetes 34}, pp. 167-191. doi: \url{10.1108/03684920510575799}.

\bibitem[Aerts \& Gabora(2005b)]{aertsgabora2005b} Aerts, D and Gabora, L. (2005b). A theory of concepts and their combinations II: A Hilbert space representation. {\it Kybernetes 34}, pp. 192-221. doi: \url{10.1108/03684920510575807}.


\bibitem[Aerts \& Sassoli de Bianchi(2015)]{aertssassolidebianchi2015} Aerts, D. and Sassoli de Bianchi, M. (2015). The unreasonable success of quantum probability II: Quantum measurements as universal measurements. {\it Journal of Mathematical Psychology 67}, pp. 76-90. doi: \url{10.1016/j.jmp.2015.05.001}.


\bibitem[Aerts \& Sassoli de Bianchi(2017)]{aertssassolidebianchi2017} Aerts, D. and Sassoli de Bianchi, M. (2017). 
{\it Universal Measurements: How to Free Three Birds in One Move}. Singapore: World Scientific. 


\bibitem[Aerts \& Sozzo(2011)]{aertssozzo2011} Aerts, D. and Sozzo, S. (2011). Quantum structure in cognition: Why and how concepts are entangled. {\it Quantum Interaction QI2011. Lecture Notes in Computer Science 7052}, pp. 116-127. doi: \url{10.1007/978-3-642-24971-6\_12}. 

\bibitem[Aerts \& Sozzo(2014)]{aertssozzo2014} Aerts, D. and Sozzo, S. (2014). Quantum entanglement in concept combinations. {\it International Journal of Theoretical Physics 53}, pp. 3587–3603. doi: \url{10.1007/s10773-013-1946-z}.

\bibitem[Aerts Argu\"elles(2018)]{aertsarguelles2018} Aerts Argu\"elles, J. (2018). The heart of an image: Quantum superposition and entanglement in visual perception. {\it Foundations of Science 23}, pp. 757-778. doi: \url{10.1007/s10699-018-9547-1}.

\bibitem[Aerts Argu\"elles \& Sozzo(2020)]{aertsarguellessozzo2020} Aerts Argu\"elles, J. and Sozzo, S. (2020). How images combine meaning. Quantum entanglement in visual perception. {\it Soft Computing 24}, pp. 10277-10286. doi: \url{10.1007/s00500-020-04692-3}. 

\bibitem[Berlin \& Kay(1969)]{berlinkay1969} Berlin, B. and Kay, P. (1969). {\it Basic Color Terms: Their Universality and Evolution}. Berkeley: University of California Press.

\bibitem[Blutner \& beim Graben(2016)]{blutnerbeimgraben2016} Blutner, R. and beim Graben, P. (2016). Quantum cognition and bounded rationality. {\it Synthese 193}, pp. 3239-3291. doi: \url{10.1007/s11229-015-0928-5}.

  
\bibitem[Bruza \& Gabora(2009)]{bruzagabora2009} Bruza, P. and Gabora, L. (Eds.) (2009). Special Issue: Quantum Cognition. {\it Journal of Mathematical Psychology 53}, pp. 303-452. doi: \url{10.1016/j.jmp.2009.06.002}.

\bibitem[Busemeyer \& Bruza(2012)]{busemeyerbruza2012} Busemeyer, J. and Bruza, P. (2012). {\it Quantum Models of Cognition and Decision}. Cambridge: Cambridge University Press.

\bibitem[Busemeyer et al.(2019)]{busemeyeretal2019} Busemeyer, J., Kvam, P. and Pleskac, T. (2019). Markov versus quantum dynamic models of belief change during evidence monitoring. {\it Scientific Reports 9}, 18025. doi: \url{10.1038/s41598-019-54383-9}.

\bibitem[Busemeyer et al.(2006)]{busemeyeretal2006} Busemeyer, J. R., Wang, Z. and Townsend, J. T. (2006). Quantum dynamics of human decision making. {\it Journal of Mathematical Psychology 50}, pp. 220-241. doi: \url{10.1016/j.jmp.2006.01.003}.

\bibitem[Collier at al.(1973)]{collieretal1973} Collier, G. A., Berlin, B. and Kay, P. (1973). Basic color terms: Their universality and evolution. {\it Language 49}, pp. 245-248. doi: \url{10.2307/412128. ISSN 0097-8507}.

\bibitem[Dalla Chiara et al.(2015)]{dallachiaraetal2015} Dalla Chiara, M. L., Giuntini, R., Leporini, R., Negri, E. and Sergioli, G. (2015). Quantum information, cognition, and music. {\it Frontiers in Psychology  6}, 1583. doi: \url{10.3389/fpsyg.2015.01583}.

\bibitem[Davidoff(2001)]{davidoff2001} Davidoff, J. (2001). Language and perceptual categorisation. {\it Trends in Cognitive Sciences 5} pp. 382-387. doi: \url{10.1016/s1364-6613(00)01726-5}.

\bibitem[Davies et al.(1998)]{daviesetal1998} Davies, I. R. L., Sowden, P. T., Jerrett, D. T., Jerrett, T. and Corbett, G. G. (1998). A cross-cultural study of English and Setswana speakers on a colour triads task: A test of the Sapir-Whorf hypothesis. {\it British Journal of Psychology. 89}, pp. 1-15. doi: \url{10.1111/j.2044-8295.1998.tb02669.x}. 


\bibitem[Disa et al.(2011)]{disaetal2011} Disa, S., LeGuen, O. and Haun, D. (2011). Categorical perception of emotional facial expressions does not require lexical categories. {\it Emotion 11}, pp. 1479-1483. doi: \url{10.1037/a0025336}.

\bibitem[Eimas et al.(1971)]{eimasetal1971} Eimas, P. D., Siqueland, E. R., Jusczyk, P. W. and Vigorito, J. (1971). Speech perception in infants. {\it Science 171}, pp. 303-306. doi: \url{10.1126/science.171.3968.303}.

\bibitem[Fubini, 1904]{fubini1904} Fubini, G., (1904). Sulle metriche definite da una forma Hermitiana. {\it Atti del Reale Istituto Veneto di Scienze, Lettere ed Arti 63} pp. 501–513. 

\bibitem[Geeraerts et al.(2001)]{geeraerts2001} Geeraerts, D., Dirven, R., Taylor, J. R. and Langacker, R. W. (Eds.), (2001). {\it Applied Cognitive Linguistics, II, Language Pedagogy}. doi: \url{10.1515/9783110866254}. 

\bibitem[Goldstone \& Hendrickson(2010)]{goldstonehendrickson2010} Goldstone, R.L. and Hendrickson, A.T. (2010). Categorical perception. {\it Wires Cognitive Science 1}, pp. 69–78. doi: \url{10.1002/wcs.26}. 


\bibitem[Harnad(1987)]{harnad1987} Harnad, S. (Ed.). (1987). {\it Categorical Perception: The Groundwork of Cognition}. Cambridge, UK: Cambridge University Press. 

\bibitem[Haven \& Khrennikov(2013)]{havenkhrennikov2013} Haven, E. and Khrennikov, A. (2013). {\it Quantum Social Science}. Cambridge:  Cambridge University Press.

\bibitem[Johanden \& Kruschke(2005)]{johansenkruschke2005} Johansen, M. K. and Kruschke, J. K. (2005). Category representation for classification and feature inference. {\it Journal of Experimental Psychology: Learning, Memory, and Cognition 31}, pp. 1433-1458. doi: \url{10.1037/0278-7393.31.6.1433}.

\bibitem[Havy \& Waxman(2016)]{havywaxman2016} Havy, M. and Waxman, S. R. (2016). Naming influences 9-month-olds’ identification of discrete categories along a perceptual continuum. {\it Cognition 156}, pp. 41-51. doi: \url{10.1016/j.cognition.2016.07.011}.

\bibitem[Hess et al.(2009)]{hessetal2009} Hess, U., Adams, R. and Kleck, R (2009). The categorical perception of emotions and traits. {\it Social Cognition 27}, pp. 320-326. doi: \url{10.1521/soco.2009.27.2.320}.

\bibitem[Khrennikov(2014)]{khrennikov2014} Khrennikov, A. (2014). {\it Ubiquitous Quantum Structure}. Berlin: Springer.

\bibitem[Lane(1965)]{lane1965} Lane, H. (1965). The motor theory of speech perception: A critical review. {\it Psychological Review 72}, pp. 275-309. doi: \url{10.1037/h0021986}.

\bibitem[Lawrence(1949)]{lawrence1949} Lawrence, D. H. (1949). Acquired distinctiveness of cues: I. Transfer between discriminations on the basis of familiarity with the stimulus. {\it Journal of Experimental Psychology 39}, pp. 770-784. doi: \url{10.1037/h0058097}.

\bibitem[Liberman et al.(1967)]{libermanetal1967} Liberman, A. M., Cooper, F. S., Shankweiler, D. P., Studdert-Kennedy, M. (1967). Perception of the speech code. {\it Psychological Review 74} pp. 431-461. doi:\url{10.1037/h0020279}.

\bibitem[Lewis, 1929]{lewis1929} Lewis, C., I. (1929). Mind and the World-order: Outline of a Theory of Knowledge. New York: Dover. 

\bibitem[Liberman et al.(1957)]{libermanetal1957} Liberman, A. M., Harris, K. S., Hoffman, H. S. and Griffith, B. C. (1957). The discrimination of speech sounds within and across phoneme boundaries. {\it Journal of Experimental Psychology 54}, pp. 358-368. doi:\url{10.1037/h0044417}.

\bibitem[Medin at al.(1984)]{medinetal1984} Medin, D. L., Altom, M. W. and Murphy, T. D. (1984). Given versus induced category representations: Use of prototype and exemplar information in classification. {\it Journal of Experimental Psychology: Learning, Memory, and Cognition 10}, pp. 333-352. doi: \url{10.1037/0278-7393.10.3.333}. 

\bibitem[Mervis et al.(1975)]{mervisetal1975}Mervis, B. C., Catlin, J. and Rosch, E. (1975).
Development of the structure of color categories. {\it Developmental Psychology 2}, pp. 54-60. doi: \url{10.1037/h0076118}.

\bibitem[Moreira \& Wichert(2016)]{moreirawichert2016} Moreira, C. and Wichert, A. (2016). Quantum probabilistic models revisited: the case of disjunction effects in cognition. {\it Frontiers in Physics 4}. 26. doi: \url{10.3389/fphy.2016.00026}.

\bibitem[Osherson \& Smith(1981)]{oshersonsmith1981} Osherson, D. N. and Smith, E. E. (1981). On the adequacy of prototype theory as a theory of concepts. {\it Cognition 9} pp. 35-58. doi:\url{10.1016/0010-0277(81)90013-5}. 

\bibitem[Pika \& Mitani, 2006]{pikamitani2006} Pika, S. and Mitani, J. (2006). Referential gestural communication in wild chimpanzees (Pan
troglodytes), Current Biology 16, R191-R192.

\bibitem[Pothos et al.(2015)]{pothosetal2015} Pothos, E. M., Barque-Duran, A., Yearsley, J. M., Trueblood, J. S., Busemeyer, J. R. and Hampton, J. A. (2015). Progress and current challenges with the quantum similarity model. {\it Frontiers in Psychology 6}, 205. doi: \url{10.3389/fpsyg.2015.00205}.

\bibitem[Regier \& Kay(2009)]{regierkay2009} Regier, T. and Kay, P. (2009). Language, thought, and color: Whorf was half right. {\it Trends in Cognitive Sciences 13}, pp. 439-447. doi: \url{10.1016/j.tics.2009.07.001}.

\bibitem[Rosch(1973)]{rosch1973} Rosch, E. H. (1973). Natural categories. {\it Cognitive Psychology 4}, pp. 328-350. doi: \url{10.1016/0010-0285(73)90017-0}. 

\bibitem[Rosch Heider(1971)]{roschheider1971} Rosch Heider, E. (1971). ``Focal'' Color areas and the development of color names.  {\it Developmental Psychology 4}, pp. 447-455. doi: \url{10.1037/h0030955}.

\bibitem[Rosch Heider(1972)]{roschheider1972} 
Rosch Heider, E. (1972). Universals in color naming and memory. {\it Journal of Experimental Psychology 93}, pp. 10-20. doi: \url{10.1037/h0032606}.


\bibitem[Rosch(1975)]{rosch1975} Rosch, E. (1975). Cognitive representations of semantic categories. {\it Journal of Experimental Psychology: General 104}, pp. 192-233. doi: \url{10.1037//0096-3445.104.3.192}. doi: \url{10.1037/h0032606}.

\bibitem[Rosch et al.(1976)]{roschetal1976} Rosch, E., Mervis, C. B, Gray, W. D, Johnson, D. M and Boyes-Braem, P. (1976). Basic objects in natural categories. {\it Cognitive Psychology. 8}, pp. 382-439. doi: \url{10.1016/0010-0285(76)90013-X}.

\bibitem[Schusterman et al.(2000)]{schustermanetal2000} Schusterman, R. J., Reichmuth, C. J. and Kastak, D. (2000). How animals classify friends and foes. {\it Current Directions in Psychological Science 9}, pp. 1-6. doi: \url{10.1111/1467-8721.00047}.

\bibitem[Sidman(1994)]{sidman1994} Sidman M. (1994). {\it Equivalence Relations and Behavior: A Research Story}. Boston, MA: Authors Cooperative.

\bibitem[Smith \& Medin(1981)]{smithmedin1981} Smith, E. E. and Medin, D. L. (1981). {\it Categories and Concepts}. Cambridge MA: Harvard University Press. 

\bibitem[Study, 1905]{study1905} Study, E. (1905). K\"urzeste Wege im komplexen Gebiet. {\it Mathematische Annalen 60}, pp. 321-378.

\bibitem[Surov et al.(2019)]{surovetal2019} Surov, I. A., Pilkevich, S. V., Alodjants, A. P. and Khmelevsky, S. V. (2019). Quantum phase stability in human cognition. {\it Frontiers in Psychology 10}, 929. doi: \url{10.3389/fpsyg.2019.00929}.

\bibitem[Tomasello, 1996]{tomasello1996} Tomasello, M. (1996).The cultural roots of language. In B. M. Velichkovskii and 
D. M. Rumbaugh (Eds.), Communicating Meaning: The Evolution and Development of Language. Mahwah, NJ: L.
Erlbaum. 

\bibitem[Yearsley(2017)]{yearsley2017} Yearsley, JM. (2017). Advanced tools and concepts for quantum cognition: a tutorial. {\it Journal of Mathematical Psychology 78}, pp. 24-39. doi: \url{10.1016/j.jmp.2016.07.005}.

\end{thebibliography}
\end{document}